# Lattice dynamical study of optical modes in $Tl_2Mn_2O_7$ and $In_2Mn_2O_7$ pyrochlores

S. Brown and H.C. Gupta[*]
*Dept. of Physics, Indian Institute of Technology, Hauz Khas, New Delhi 110016, India*

J.A. Alonso and M.J. Martínez-Lope
*Instituto de Ciencia de Materiales de Madrid, Consejo Superior de Investigaciones Cientificas, Cantoblanco, E-28049, Madrid, ,Spain*

The Raman, IR and force field have been investigated for $A_2Mn_2O_7$ (A= Tl, In) by means of a short-range force constant model which includes four stretching and four bending force constants. Unusual spectral and force field changes are observed and analyzed. The stretching force constant Mn-O and A-O' are found to be relatively higher than those of other pyrochlore oxides of the $A_2Mn_2O_7$ family, while the remaining force constant values are significantly smaller, especially for $Tl_2Mn_2O_7$. This trend may be due to strong hybridization of the Tl (6s) orbital with O (2p) and Mn (3d). The assignment of all the modes has been proposed and potential energy distribution is also reported. The evaluated frequencies are close to the available observed infrared and Raman frequencies, giving further support to the present assignments.

# I. INTRODUCTION

The discovery of colossal magnetoresistance (CMR) in manganese oxides has generated much interest in both basic research and possible applications[1-6]. CMR effects in pervoskite (namely $La_{1-x}Sr_xMnO_3$) and pyrochlore (namely $Tl_2Mn_2O_7$) manganese oxides are found in the vicinity of a metal-insulator transition and a ferromagnetic ordering transition, although the origin of the CMR behavior in both families of compounds is very different. In pervoskites, the CMR behavior is explained within the framework of the double-exchange model (DE). The Jahn-Teller distortion and a strong electron–phonon coupling also play an important role in perovskites[7]. On the contrary, the pyrochlore ferromagnetism ( $T_C$= 120 K) is a consequence of super-exchange interactions between $Mn^{4+}$ magnetic moments, whereas the metallic behavior of $Tl_2Mn_2O_7$ originates from the hybridization among Tl(6s), O(2p), and Mn(3d)[8-9], which is probably at the very origin of the magneto-transport properties of this peculiar pyrochlore.

At room temperature, most $A_2B_2O_7$ pyrochlores exhibit a face centered cubic structure belonging to the space group (Fd3m, $O_h^7$), no. 227; the primitive cell contains two formula units. The pyrochlore $A_2B_2O_6O'$ structure (Fig.1) may be considered as interpenetrating networks of $B_2O_6$ and $A_2O'$; although the latter is not essential for the stability of the structure[10]. The $B_2O_6$ sublattice is formed by $BO_6$ octahedra sharing corners, with B-O-B angles close to 135º. Manganese pyrochlores of stoichiometry $A_2Mn_2O_7$ have been described for trivalent A cations such as rare earths, In and Tl. The transport and magnetic properties dramatically depend upon the nature of the A cation. All compounds exhibit ferromagnetic interactions, with Curie temperatures around 15 K (for instance, $Y_2Mn_2O_7$ or $Lu_2Mn_2O_7$), excepting $Tl_2Mn_2O_7$ and $In_2Mn_2O_7$ which exhibit much higher $T_C$'s of 120 K. Moreover, only the $Tl_2Mn_2O_7$ pyrochlore exhibit semi-metallic conductivity and CMR properties[8]. These differential properties have been explained as a result of a strong hybridization between the Tl (6s), O (2p) and Mn (3d) orbitals. It is plausible to expect that the different nature of the chemical bond in these oxides will also influence the phonon spectra, and hence a study of phonons is undertaken in the present work for $A_2Mn_2O_7$ (A= In, Tl).

In fact, very interesting information about the physics of these materials can be learned through light scattering and absorption experiments. Raman and IR spectroscopy can give insight to the dynamical processes caused by phonons, charge carriers and spins that may effect the magneto- resistance process. The force field is also one of the important physical factors, which reflects the electronic structure of the bond and its variations with the bond environment. In terms of total amount of research devoted so far to the pyrochlore materials, surprisingly little work appears to have been done on the spectroscopic studies and lattice dynamics of these compounds[11-12]. In particular, the available information about the assignment of the observation to specific modes of vibrational and force field study is still lacking. Motivated from this situation, we thought it pertinent to study the Raman, IR Spectra and force field of $A_2Mn_2O_7$ (Tl, In). The assignment of all the vibrational modes has been made. The results are also compared and discussed with other $A_2Mn_2O_7$ (A= Y, Dy, Er, Yb) and different pyrochlore oxides [13-16].

## II. EXPERIMENTAL AND THEORY

The elaboration of $In_2Mn_2O_7$ pyrochlore required the previous preparation of very reactive precursor, obtained by wet-chemistry techniques. A mixture of $In_2O_3$ and $MnCO_3$ was dissolved in citric and nitric acid; the solution was slowly evaporated, leading to an organic resin which was dried at 120°C and slowly decomposed at temperatures up to 600°C in order to eliminate all the organic materials and nitrates. This precursor powder was thoroughly ground with $KClO_4$ (30% in weight), put into a gold capsule (8mm dia., 10mm length), sealed and placed in a cylindrical graphite heater. The reaction was carried out in a piston-cylinder press at a pressure of 2 GPa at 1000°C for 60 min. Then the material was quenched to room temperature and the pressure was subsequently released. The raw product, obtained as a dense, homogeneous pellet, was ground and washed in water, in order to dissolve KCl coming from the decomposition of $KClO_4$; then the powder sample was dried in air at 150°C for 1 h. The preparation of $Tl_2Mn_2O_7$ was performed by solid state reaction from $Tl_2O_3$ and $MnO_2$, ground together and put into a gold capsule. The high pressure reaction conditions were the same described for the $In_2Mn_2O_7$ compound.

Raman measurements were done in a Raman spectrometer with an Olympus microscope, a Jobin-Yvon HR 460 monochromator and a $N_2$ cooled CCD. All spectra were obtained at room temperature, exciting with the 514.5 nm line of a Spectra Physics Ar-Kr laser and Super-Notch Plus filter (Kaiser Optical Systems) to suppress the Rayleigh scattering. The IR spectra (50-4000 cm$^{-1}$) were recorded from KBr and polyethylene pellets in IR BRUKER–IFS66v/S spectrophotometer.

The normal coordinate analysis treatment was carried out according to the Shimanaouchi method[17] based on the Wilson GF method. A short-range force field model (SRFCM) was chosen. In the GF matrix method, the eigen value equation is given by $|FG - E\lambda| = 0$, where the F matrix contains force constants that describes the vibrational potential energy of system. The G matrix involves the mass number and certain spatial relationship of the atoms, thus bringing the kinetic energies into equation. E is unit matrix of the same dimension of F and G, and $\lambda$ is defined by $\lambda = 4\pi^2 c^2 \nu^2$. The force constant matrix includes short range valence forces between nearest neighbors Mn–O, A–O', A–O; force between repulsive nearest neighbors O - O' and angle bending forces between O-Mn-O, O-Mn-O, O-A-O' and O-A-O. These eight force constants are optimized to give the best fit of observed Raman and infrared wavenumbers.

### III. RESULTS AND DISCUSSION

The group theoretical analysis for optical and acoustical modes from correlation method[16] and taking account the symmetry conditions are listed in Table 1, we have:

$\Gamma = A_{1g} + E_g + 2F_{1g} + 4F_{2g} + 3A_{2u} + 3E_u + 8F_{1u} + 4F_{2u}$

Out of these 26 normal modes, only $A_{1g}$, $E_g$, $4F_{2g}$ are Raman active, $7F_{1u}$ are infrared active and one $F_{1u}$ acoustical. All the seven $F_{1u}$ modes and six Raman active modes are observed in most cases. The Raman and infrared spectra of $A_2Mn_2O_7$ (Tl, In) are shown in Fig.2 and 3 respectively. For comparison, the spectra of other manganese pyrochlores are also displayed[16]. The $A_{1g}$ (totally symmetric mode) is readily assigned as it gives rise to a strong polarized line[12]. The observed $A_{1g}$ and $F_{2g}^1$ are least for $Tl_2Mn_2O_7$ compared to other manganese pyrochlores. The most striking feature of the $Tl_2Mn_2O_7$ spectra is that, in general, the frequencies are lower than those observed for other pyrochlores, but some frequencies are increased in $Tl_2Mn_2O_7$ like $F_{1u}^1$ and $F_{1u}^2$. The $F_{1u}^2$ and $F_{1u}^3$ modes become quite distinct as the mass increases. The lower range frequencies $F_{1u}^5$, $F_{1u}^6$, $F_{1u}^7$ are almost the same for both In and Tl compounds, though the higher range frequencies of $In_2Mn_2O_7$ are comparable to those of other manganese pyrochlores. The bands are assigned to symmetry species by comparing the previously published study on pyrochlore oxides[13-16,19]. The extra band around 700 cm$^{-1}$ is observed in all the Raman spectra presented in this work and has been assigned as an overtone[20-21].

The eight internal coordinate sets introduced in the force field calculations and their definition is given in Table 2. The refined force constants are also reported in Table 2. The comparison between the observed and the calculated wavenumbers are given in Table 3. A satisfactory fit (average error of 5% between observed and calculated wavenumbers) was obtained with four stretching and four bending force constants. As the octahedra Mn-$O_6$ is the backbone of the pyrochlore structure, a similar trend is observed in calculated force constants. The most dominating force constants belonging to the octahedra are found to be K1, H5 and H6. K1 and K2 values are the highest for $Tl_2Mn_2O_7$ among all the pyrochlore oxides till now studied. K1, K2 and H6 force constant increases as A change from RE (Y, Dy, Er, Yb) to Tl whereas the other force constants decrease. This trend may be due to the hybridization among Tl(6s), O(2p) and Mn(3d) orbitals, accounting for an strengthening of the Mn-O and Tl-O chemical bonds. For other manganese pyrochlores, a very slight change is observed in these force constants from Y to Yb pyrochlores. Although the experimental Raman and infrared wavenumbers also do not exhibit any systematic variation when A changes from Y to Yb, a general trend is observed along the series, leading to a decrease in the force constant with an increase in bond length.

Comparing the obtained force constants from the earlier studied pyrochlore oxides[12, 16-17], it has been observed that in $BO_6$ octahedra (B=Mn, Ti, Hf, Sn), Mn cations are more tightly bound to oxygens than Ti, Hf and Sn in titanates, hafnates and stannates. In manganates $A_2Mn_2O_7$, the stretching force constant (K1) has the largest value observed for any pyrochlore oxides.

The potential energy distribution (PED), which gives important information to understand the energy localization of the active modes, has a similar trend for both Tl and In compounds as shown in Fig 4 and 5. From PED, it is observed that $A_{1g}$ and $F_{2g}^1$ modes are due to the bending of Mn-$O_6$ octahedra. Similarly, $F_{2g}^2$ and $F_{2g}^3$ are mainly due to the stretching of Mn-O and A-O' bonds respectively. Hence the increase in frequency can be correlated well with the increase in stretching force constants K1 and K2. From PED, it is obvious that $F_{1u}^1$ has the most dominating force constant K1 and as the measured frequency is highest among all the manganates, the highest value of K1 for $Tl_2Mn_2O_7$ is justified. $F_{1u}^2$ and $F_{1u}^3$ have important contribution from H5 and K2 respectively. Also $F_{1u}^5$ is due to stretching vibration of A-$MnO_6$ and $F_{1u}^6$ & $F_{1u}^7$ are mainly influenced by the bending vibrations of $AO_6O'$ polyhedra. This is the reason behind why $F_{1u}^6$ and $F_{1u}^7$ frequencies differ more than other frequencies when A changes from Y to lanthanide elements. For mangantes pyrochlores, it is also observed that the three bending force constants have important contribution in PED (H5, H6, and H8) but earlier studied titanates, hafnates and stannates do not display much contribution from H8.

## IV. CONCLUSION

We have applied the SRFC model involving various stretching and bending force constants to evaluate the Raman and IR wavenumbers in the case of $Tl_2Mn_2O_7$ and $In_2Mn_2O_7$. The force constant values suggest that due to hybridization of different states Tl(6s), O(2p), and Mn(3d) force constants K1, K2 and H5 dominate in the force field while others decrease. The calculated IR and Raman modes exhibit a satisfactory agreement with the experimental observation and for the first time, an assignment of these frequencies to species modes has been established.

## ACKNOWLEDGMENTS

S.B. is grateful to Dr. J.A. Alonso and co-workers for providing hospitality at ICMM, Madrid, where most of the work was completed, and also to Dr. A. de Andrés and J. Sánchez-Benítez (ICMM) for their assistance with the Raman measurements. Financial support from the Spanish Ministry of Science and Technology, project MAT2001-0539 and Senior Research fellowship from CSIR (India) is gratefully acknowledged.

TABLE I. Factor group analysis for the zone-center vibrational modes of pyrochlore $A_2B_2O_6O'$.

| Ion | Number of equivalent positions (Wyckoff notation) | Site Group Symmetry | Irreducible Representations |
|---|---|---|---|
| A | 16(c) | $D_{3d}$ | $A_{2u} \oplus E_u \oplus 2F_{1u} \oplus F_{2u}$ |
| B | 16(d) | $D_{3d}$ | $A_{2u} \oplus E_u \oplus 2F_{1u} \oplus F_{2u}$ |
| O | 48(f) | $C_{2v}$ | $A_{1g} \oplus E_g \oplus 2F_{1g} \oplus 3F_{2g} \oplus A_{2u} \oplus E_u \oplus 3F_{1u} \oplus 2F_{2u}$ |
| O' | 8(a) | $T_d$ | $F_{1u} \oplus F_{2g}$ |
| | Total | | $\Gamma = A_{1g} \oplus E_g \oplus 2F_{1g} \oplus 4F_{2g} \oplus 3A_{2u} \oplus 3E_u \oplus 8F_{1u} \oplus 4F_{2u}$ |
| | Acoustic | | $\Gamma_{ac} = F_{1u}$ |
| | Raman | | $\Gamma_R = A_{1g} \oplus E_g \oplus 4F_{2g}$ |
| | Infrared | | $\Gamma_{IR} = 7 F_{1u}$ |

TABLE II. Definition of internal coordinates and force field (in N.cm$^{-1}$) for $A_2Mn_2O_7$ (A= Y, In, Dy, Er, Yb, Tl). The values for Y, Dy, Er and Yb have been taken from Ref. [16].

| Internal Coordinates | | Force constant (N.cm$^{-1}$) | | | | | |
|---|---|---|---|---|---|---|---|
| | | $Y_2Mn_2O_7$ | $Dy_2Mn_2O_7$ | $Er_2Mn_2O_7$ | $Yb_2Mn_2O_7$ | $In_2Mn_2O_7$ | $Tl_2Mn_2O_7$ |
| K1 | Mn-O | 0.776 | 0.878 | 0.787 | 0.734 | 0.843 | 1.027 |
| K2 | A-O' | 0.227 | 0.317 | 0.348 | 0.306 | 0.369 | 0.873 |
| K3 | A-O | 0.096 | 0.186 | 0.192 | 0.194 | 0.234 | 0.088 |
| K4 | O-O' | - | - | - | - | 0.045 | 0.026 |
| H5 | O-Mn-O | 0.274 | 0.276 | 0.282 | 0.263 | 0.331 | 0.168 |
| H6 | O-Mn-O | 0.376 | 0.343 | 0.361 | 0.383 | 0.248 | 0.402 |
| H7 | O-A-O | 0.073 | 0.031 | 0.046 | 0.074 | - | 0.05 |
| H8 | O-A-O' | 0.246 | 0.212 | 0.208 | 0.208 | 0.148 | 0.09 |

TABLE III. Comparison of calculated and experimental values of phonon wavenumbers (in cm$^{-1}$) near the Brillouin zone center at room temperature.

|         | In$_2$Mn$_2$O$_7$ |           | Tl$_2$Mn$_2$O$_7$ |           |
|---------|-------------------|-----------|-------------------|-----------|
|         | Calculated        | Observed* | Calculated        | Observed* |
| $A_{1g}$    | 511.3 | 510 | 490.9 | 489 |
| $E_g$       | 344.1 | 346 | 317.3 | 327 |
| $F_{2g}^1$  | 547.1 | 548 | 523.5 | 512 |
| $F_{2g}^2$  | 442.1 | 428 | 495.0 | -   |
| $F_{2g}^3$  | 365.1 | -   | 412.2 | 392 |
| $F_{2g}^4$  | 303.0 | 292 | 296.6 | 289 |
| $F_{1u}^1$  | 535.7 | 542 | 589.0 | 595 |
| $F_{1u}^2$  | 483.5 | 482 | 492.8 | 512 |
| $F_{1u}^3$  | 436.9 | -   | 436.1 | 454 |
| $F_{1u}^4$  | 347.4 | 336 | 389.1 | -   |
| $F_{1u}^5$  | 266.9 | 280 | 288.3 | 275 |
| $F_{1u}^6$  | 144.0 | 140 | 123.1 | 125 |
| $F_{1u}^7$  | 131.  | -   | 97.0  | 98  |

*This work.

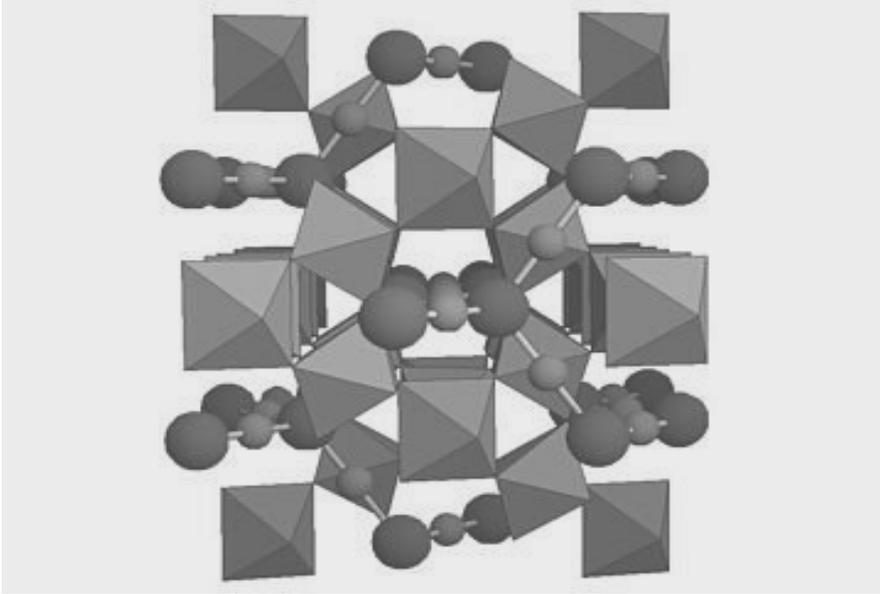

FIG. 1. $A_2B_2O_7$ cubic pyrochlore structure with $BO_6$ octahedra and $A_2O'$ chains.

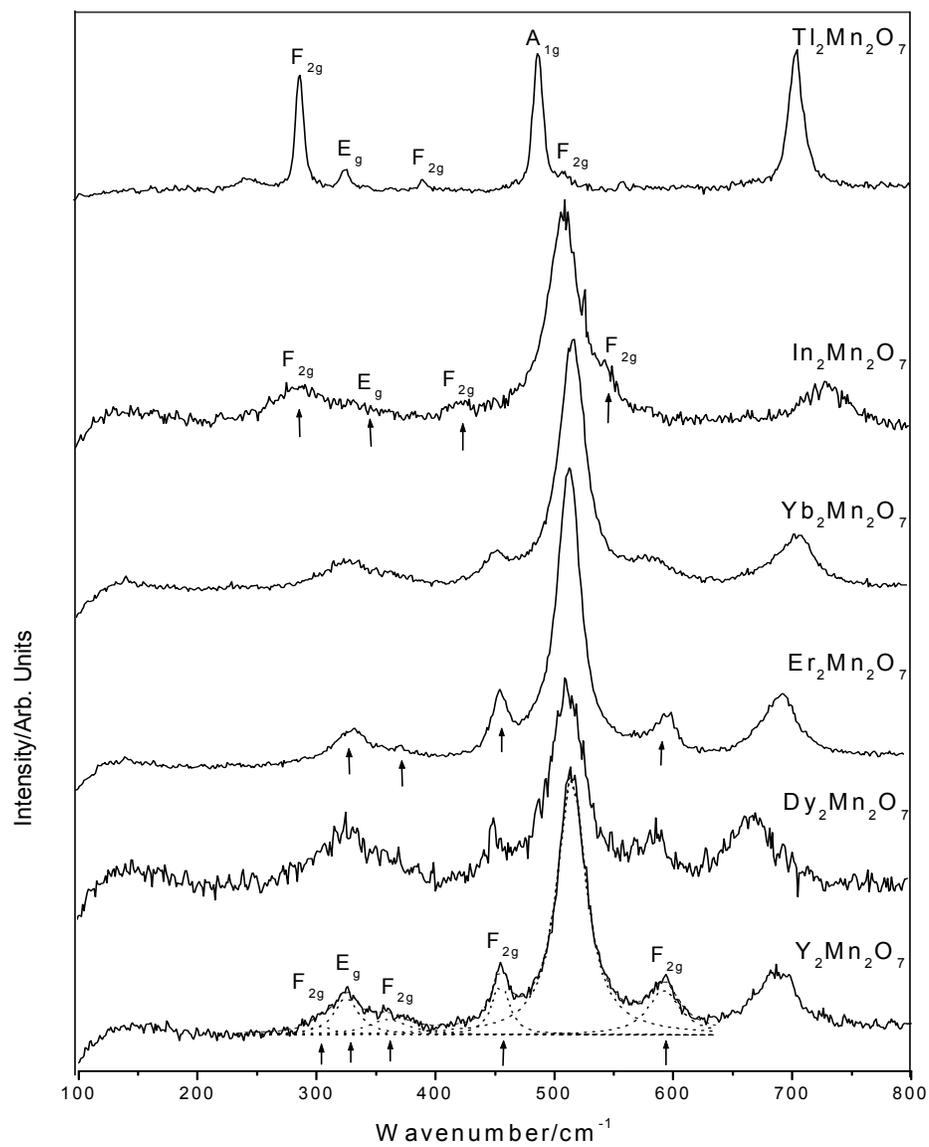

FIG. 2. Raman spectra of $A_2Mn_2O_7$ (A= Y, In, Dy, Er, Yb, Tl) pyrochlore.

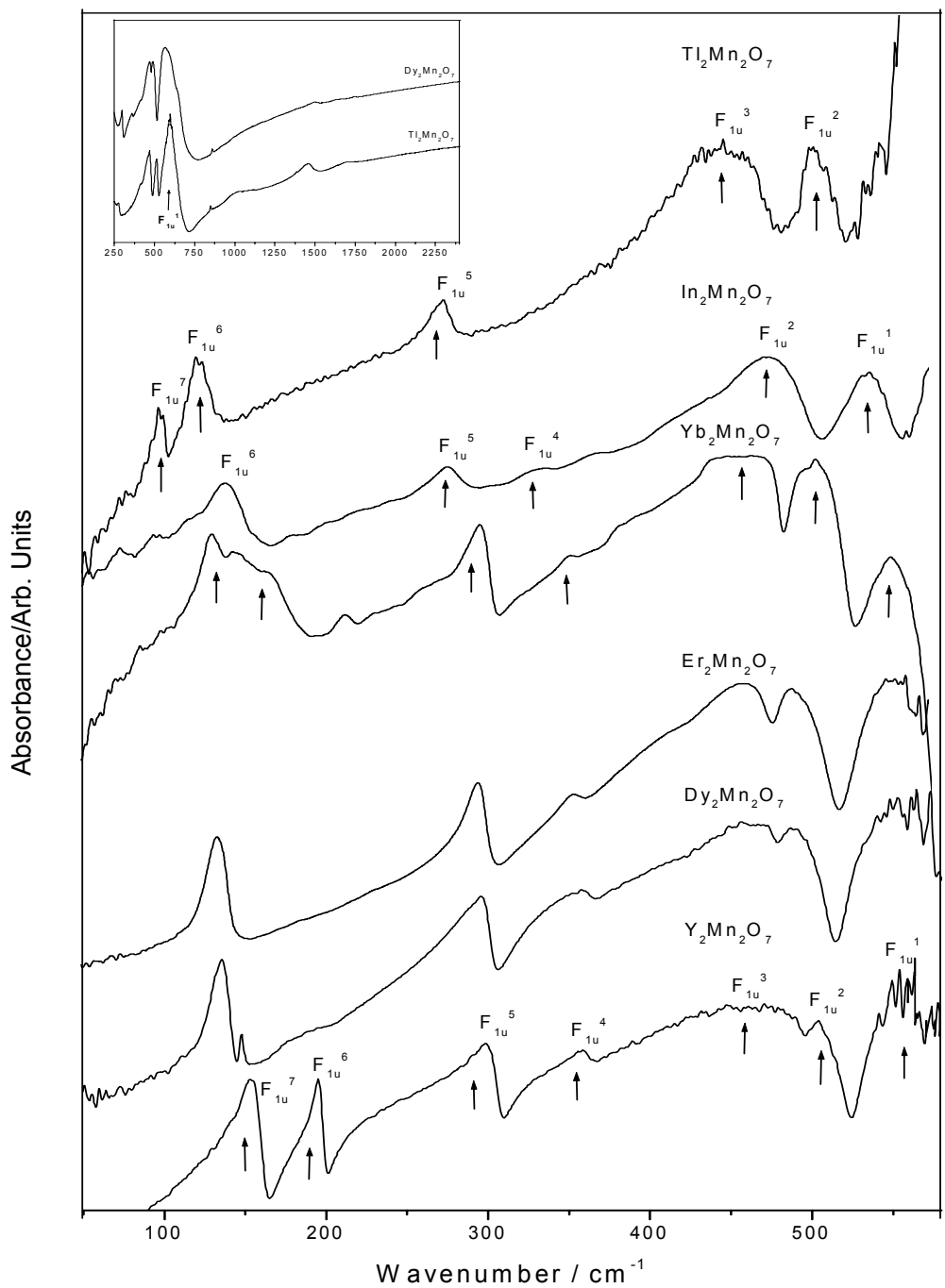

FIG. 3. Infrared spectra of the $A_2Mn_2O_7$ (A=Y, Dy, Er, Yb, In, Tl) pyrochlores. The inset shows Far IR spectra of $A_2Mn_2O_7$ (A= Dy, Tl).

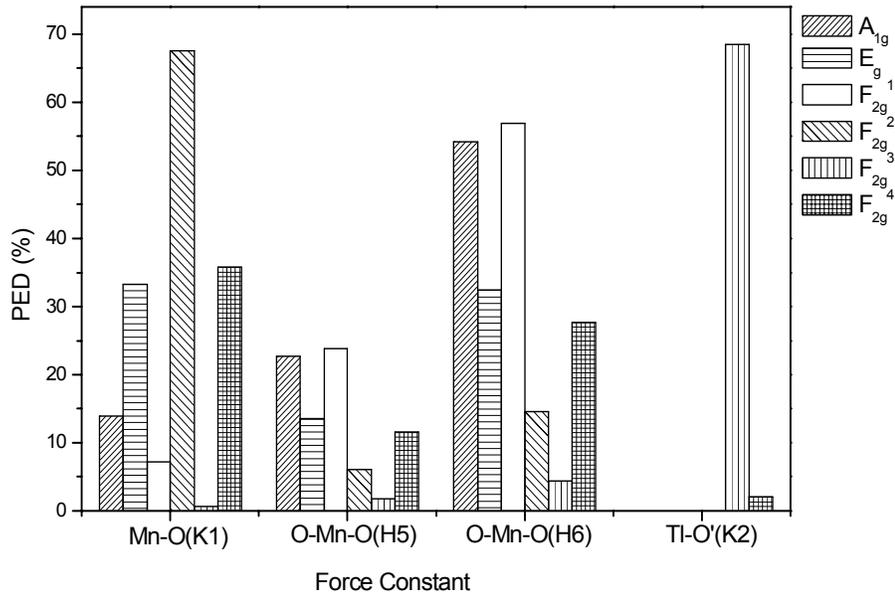

FIG. 4. Potential energy distribution (PED) of Raman active modes for $Tl_2Mn_2O_7$.

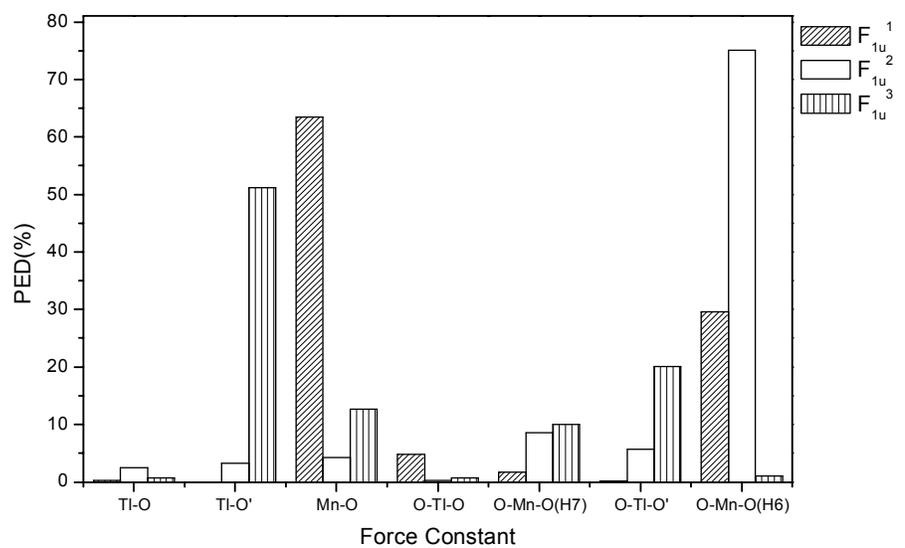

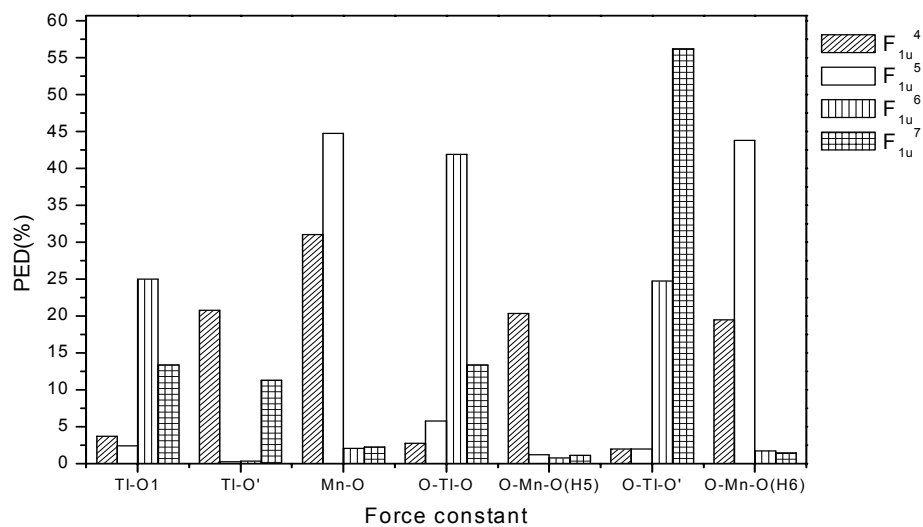

FIG. 5. Potential energy distribution (PED) of IR active modes for $Tl_2Mn_2O_7$.